\documentclass{ws-ijmpa}

\begin{document}

\markboth{V.~A.~Beylin, V.~I.~Kuksa, R.~S.~Pasechnik,
G.~M.~Vereshkov} {Neutralino-nucleon interaction in the Split SUSY
scenario of the Dark Matter}

%
\catchline{}{}{}{}{}
%

\title{NEUTRALINO-NUCLEON INTERACTION IN THE SPLIT SUSY\\ SCENARIO OF
THE DARK MATTER}

\author{R.~S.~PASECHNIK}

\address{Bogoliubov Laboratory of Theoretical Physics, JINR,
Dubna 141980, Russia}

\author{V.~A.~BEYLIN, V.~I.~KUKSA, G.~M.~VERESHKOV}

\address{Institute of Physics, Southern Federal University, Rostov-on-Don 344090, Russia}

\maketitle

\begin{history}
\received{Day Month Year} \revised{Day Month Year}
\end{history}

\begin{abstract}
The Split SUSY scenario with light Higgsino states is treated as an
application to the Dark Matter problem. We have considered the
structure of the neutralino-nucleon interaction and calculated
cross-section of the neutralino-nucleon scattering. The decay
properties of the lightest chargino and next lightest neutralino are
analyzed in details.

\end{abstract}

\ccode{PACS numbers: 12.60.Jv, 95.35.+d, 95.30.Cq}

\section{Introduction}

One of the most important phenomenological applications of the
Supersymmetry (SUSY) is the treatment of neutralino as a candidate
for the (Cold) Dark Matter (DM). In Refs.~[\refcite{1,1a,2,2a,3}]
new hierarchies of SUSY scales (so called Split Supersymmetry
scenarios) were motivated, on the one hand, by the Anthropic
Principle and, on the other hand, multi-vacua string landscape
arguments. Scenarios of this type are alternative to the MSSM in
some aspects and do not reject the fine-tuning mechanism in the
spectrum of scales [\refcite{4,5,6,6a,7,8,9}].

The Renormalization Group (RG) analysis of the Split SUSY models was
performed in Refs.~[\refcite{2,3,4,8,9}]. In Ref.~[\refcite{9}] the
one-loop RG behavior of SUSY $SU(5)$ was considered in detail
accounting for degrees of freedom in the vicinity of $M_{GUT}$
scale. It was noticed that these heavy states should be considered
as threshold corrections; they are important for the final
arrangement of scales providing sufficiently high unification point.
From the RG consideration at the one-loop level two classes of
scenarios emerge having an opposite arrangement of $\mu$ and
$M_{1/2}$ scales ($M_{1/2}$ is an order of $M_1$ or $M_2$). At the
same time, the analysis does not fix the characteristic scale of
superscalars, $M_0$, due to a specific form of the one-loop
equations $\mu,M_{1/2}=f_a(M_{GUT})$ following from the RG analysis.
More precisely, these equations contain the squark and slepton
scales in ratios $M_{\tilde q}/M_{\tilde l}$ only.

For the first class of scenarios the hierarchy $|\mu|\gg M_{1/2}$
takes place. The second class is defined by the opposite hierarchy
$|\mu|\ll M_{1/2}$. In particular, the RG analysis [\refcite{9}]
results in the hierarchy
\begin{equation}\label{1.1}
 \displaystyle M_{0}\sim
   M_{1,2}\gg |\mu| > M_{EW},
\end{equation}
whose spectrum contains two lightest neutralinos degenerated in mass
(almost pure Higgsino case) and one of charginos as the nearest to
the electroweak scale $M_{EW}$. In this scenario both $M_0$ and
$M_{1/2}$ can be shifted to scales $\sim (10^6 - 10^{10}) \,
\mathrm{GeV}$. As it was shown in Ref.~[\refcite{7,9}], in one-loop
approximation the RG does not fix these scales strictly due to
uncertainties in dimensionless parameters, which are defined by the
heavy scales ratios. Usually assumed value for this scale is $\sim
10^8 - 10^9  \mathrm{GeV}$ and it is within our RG motivated
interval.

In this paper, we study some particular features of the scenario
(\ref{1.1}) and their consequences in experiments. We consider the
basic characteristics of the neutralino manifestations -- the
cross-section of the neutralino-nucleon scattering and decay
properties of the neutralino and chargino. We will explicitly show
that the direct observation of relic neutralino (Higgsino) in the
$\chi-N$ scattering is impossible due to the fact that the typical
energy of relic $\chi$ is far below the corresponding threshold (see
Section 2). So, we concentrate our attention on the high-energy
neutralino-nucleon scattering and decays of its products.

The program of collider experiments is mainly based on study of the
MSSM, which has supersymmetric degrees of freedom near
$1\,\mbox{TeV}$. At this scale the Split Supersymmetry displays the
isolated lightest scales only. In particular, in the scenario under
consideration there are two lightest neutralino (LSP, $\chi_1^0$ and
NLSP, $\chi_2^0$) and one light chargino $\tilde{H}$ with the masses
$\sim1\,\mbox{TeV}$.

The signature of the neutralino and chargino production and decays
at the LHC was considered in many papers. It was shown that this
signature crucially depends on values of the mass splitting
[\refcite{8}], [\refcite{10}]-[\refcite{13f}]. In these papers main
attention was paid to calculation of the cross-section of neutralino
and chargino production at hadron colliders. The radiative decays at
the loop level were considered in Ref.~[\refcite{11}]. Decay rates
in semileptonic and hadronic (production of quarks and jets) decay
channels of neutralino and chargino were discussed there in some
detail. It was pointed out that, on the one hand, observation of
these hardly detectable effects would mean the possibility to gain
important information about the scales of higher SUSY states (in
particular, $\tilde t_{1,2}$ and their mixing) just from the
one-loop mass splitting
calculations~[\refcite{14-1}]-[\refcite{16}]. On the other hand, if
only low-lying neutralino and chargino are detected in experiments
near TeV scale, the conventional MSSM spectrum should be necessarily
split in some manner to produce higher scales for other superstates.
The spectrum of the lowest states, $\chi_{1,\,2}$ (LSP and NLSP) and
$\tilde{H}$ (chargino), is nearly degenerated. For the scenario
(\ref{1.1}) this fact is well known (see, for example,
Refs.~[\refcite{16}]-[\refcite{19}]). Two other neutralino states,
$\chi_{3,\,4}$, and heavy chargino $\tilde{W}$ are placed far from
the lightest ones at the scale $\sim M_{SUSY}$.

According to the well-known method and results of the relic
abundance analysis (Refs.~[\refcite{20}]-[\refcite{25}] and
references therein) neutralinos before their freeze-out live in the
thermodynamical equilibrium with other components of the
cosmological plasma. In order to compare the calculated value of
relic abundance $\Omega h^2$ with the corresponding experimental
corridor given by the relic data [\refcite{26}], we have used the
known values of SUSY parameters and extracted the following LSP
(Higgsino) mass [\refcite{7,9}]: $M_{\chi}= 1.0 - 1.4
\;\mathrm{TeV}$ for $x_f=25$ and $M_{\chi} = 1.4 - 1.6
\;\mathrm{TeV}$ for $x_f=20$. These values do not break the gauge
coupling convergence and are in good agreement with the results of
Refs.~[\refcite{2,2a,3,8,10}]. Thus, in the model where these two
lightest neutralinos and one chargino are closest to the EW scale,
they have masses $O(1\;\mathrm{TeV})$. Further, we will use
$M_{\chi} = 1.4 \;\mathrm{TeV}$ as an average value for all
numerical estimations.

In this paper, we consider the possibility of registration of
neutralino-nucleon scattering, when neutralinos are low-energy
(relic) and high-energy (non-relic) ones. We show that in the first
case the process is closed in the framework of the scenario under
consideration. We calculate cross-section of the high-energy
neutralino-nucleon scattering with the production of the lightest
chargino and NLSP. We also consider in detail the decay rates of
these products for the kinematically allowed channels.

The structure of the paper is as follows. In Section 2, the
cross-section of neutralino-nucleon scattering is considered for
low- and high-energy neutralino. The decay properties of the
lightest chargino $\tilde{H}$ and NLSP $\chi_2$ are analyzed in
Section 3 and 4. In these sections we describe the results of
calculations, which are necessary for discussion of possible
experimental manifestations of the considering SUSY scenario.
Finally, some conclusions are given in Section 5. Appendices A and B contain an
important details needed for calculations.

\section{The neutralino-nucleon scattering}

Supposing that the lightest neutralino $\chi_1$ is the main DM
constituent, experimental manifestations of the Split Higgsino
scenario crucially depend on the neutralino mass splitting
parameters $\delta m=M_{\chi_2} - M_{\chi_1}$ and $\delta m^- =
M_{\tilde{H}}- M_{\chi_1}$. These mass splittings are determined by
the sum of their tree values and radiative corrections. Tree level
mass splitting is approximately defined as [\refcite{11,17}]:
\begin{equation}\label{3}
\delta m\approx M^2_Z/M,\,\,\,\delta m\approx 2\delta m^-,
\end{equation}
where $M\sim M_{1,2}$ and we suppose $\tan\beta\gg1$. So, in the
interval $M\sim(10^6-10^{10})\,\,\,\mbox{GeV}$ the splitting is
rather small: $\delta m\sim (0.001-10)\,\,\,\mbox{MeV}$. In the
scenario under consideration, one-loop diagrams with $\gamma, Z$ and
$W$ bosons in the intermediate state contribute mainly to the value
$\delta m^-$. Calculations of this contribution were performed in
Ref.~[\refcite{7,11}] in various ways, and the same result was
obtained, $\delta m^- \approx 350\,\,\,\mbox{MeV}$. To evaluate the
contribution of the gauge bosons to $\delta m$, we should take into
account the diagonal and non-diagonal self-energy contributions to
the mass matrix [\refcite{19}]. However, this problem was not
considered in detail for our case. It is usually assumed that loop
corrections from the heavy $\tilde t_a$ and $\tilde b_a, \, a=1,2$
states (they are all at the high $M_0$ scale) to the mass splittings
are negligible in such ``low$-\mu$'' scenarios. Note, however, that
condition of the splitting smallness depends on the structure of
these high energy states. If there is a significant gap in the
superscalars mass spectrum, loop corrections both to $\delta m$ and
$\delta m^-$ can be comparable with their tree values or even exceed
them~[\refcite{12-1,12-2,14-1,14-2,14a,14b,15,16,18,19}]. As it is
known [\refcite{16}], the hierarchy of $\tilde t_1$ and $\tilde t_2$
states and their mixing angle $\theta_t$ drive the value of the mass
difference when squarks dominate in loops (this simple approximation
takes place when $m_{\tilde 1}^2\gg m_{\tilde 2}^2$):
\begin{equation}\label{3.01}
 \delta m\approx 2G^2_t m_t\sin(2\theta_t)\cdot
\ln(\frac{m^2_{\tilde{t}_1}}{m^2_{\tilde{t}_2}}),
\end{equation}
where
\begin{equation}
G_t=\sqrt{\frac{3G_F}{8\sqrt{2}\pi^2}}\frac{m_t}{\sin\beta}.
\end{equation}
For $\sin\beta\approx1$ and $\sin 2\theta_t\approx1$, from
Eq.~(\ref{3.01}) it follows $\delta m \approx 5
\log(m^2_{\tilde{t}_1}/m^2_{\tilde{t}_2})\,\,\,\mbox{GeV}$. So,
despite of heaviness of scalars in this scenario, their contribution
to the mass splittings can be large if, for example, the ratio
$m_{\tilde t_1}/ m_{\tilde t_2} \gtrsim 3$. Because of lack of
information about the hierarchy of the squark masses, mixing angle
$\theta_t$ and some details of loop contributions to the mass
splittings within the Split SUSY scenarios, we assume $\delta m,
\,\,\delta m^-\lesssim 1-2\,\,\mbox{GeV}$. Then, we will consider
the decay properties of $\tilde{H}$ and $\chi_2$ in Sections 3 and 4
for these values of mass splittings. At any rate, this consideration
(together with the analysis of possible experimental data on NLSP
and chargino creation and decays) can provide some information on
higher scale states in the split mass spectrum.

Here we discuss some theoretical possibilities for the direct
detection experiments that can be given by the neutralino-nucleon
interactions. This interaction structure strongly depends on the
structure of the neutralino-boson interaction. Just this point
demands very accurate mathematical analysis, which was performed in
Ref.~[\refcite{17}] (see also Appendix A). In our calculations we
use $L_{int}$ in the form [\refcite{17}]:
\begin{eqnarray}\nonumber
 L_{int}&=&g_2W^+_{\mu}\left(-\frac{i}{2}\bar{\chi}_2\gamma^{\mu}\tilde{H}
          -\frac{1}{2}\bar{\chi}_1\gamma^{\mu}\tilde{H}\right)
        +g_2W^-_{\mu}\left(+\frac{i}{2}\bar{\tilde{H}}\gamma^{\mu}\chi_2
          -\frac{1}{2}\bar{\tilde{H}}\gamma^{\mu}\chi_1\right)\\
        &+&\frac{ig_2}{2\cos{\theta_W}}Z_{\mu}\bar{\chi}_2\gamma^{\mu}\chi_1,
 \label{3.0}
\end{eqnarray}
where the only light states are taken into account. Note, the
important feature of the $Z\chi\chi$ vertex which follows from the
(\ref{3.0}): the dominant contribution to this vertex is given by
the non-diagonal vector-like term
$Z_{\mu}\bar{\chi}_2\gamma^{\mu}\chi_1$, while the axial vector
terms $Z_{\mu}\bar{\chi}_i\gamma^{\mu}\gamma_5\chi_k$ are suppressed
by the small mixing with heavy neutralino states. This fact directly
follows from the connection between the neutralino mass sign, parity
and structure of the neutralino-boson interaction [\refcite{17}].
Such interaction structure leads to the dominant contribution to the
spin-independent (SI) part of the neutralino-nucleon scattering in
non-relativistic region (see later, Eq. (\ref{3.1}) and Appendices A
and B).

\begin{figure}[!h]
\centerline{\epsfig{file=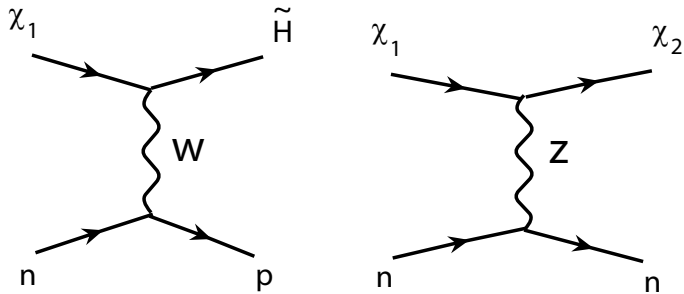,width=9cm}} \caption{Feynman
diagrams for neutralino-nucleon scattering.} \label{fig:Feynm1}
\end{figure}

Processes of the lightest neutralino-nucleon scattering with
$\chi_2$ or $\tilde{H}$ in the final state are presented in
Fig.~\ref{fig:Feynm1}.

We consider two different cases of the process -- the scattering of
relic non-relativistic neutralino and the scattering of high-energy
neutralino which can be produced by cosmic rays or in decays of an
exotic $M_X$ particles (for a review of such an option see, for
example, [\refcite{27,28,29}]).

Neutralino-boson interaction in the considering scenario is
described by the corresponding term in the Lagrangian (\ref{3.0}).
Effective low-energy interaction of nucleons and bosons is described
by the vertex $(ig/4\cos\theta_W)\gamma_{\mu}(c_V-c_A\gamma_5)$.
Amplitude of the scattering is given by standard calculation rules:
\begin{equation}\label{3.a}
\mathit{M}=\frac{ig^2}{8\cos^2\theta_W M^2_Z} \bar{\chi}_2\gamma^{\mu}\chi_1\cdot \bar{N}\gamma_{\mu}(c_V-c_A\gamma_5)N,
\end{equation}
where $N=n,p$, $c_V=1$ for neutron, $c_V=1-4\sin^2\theta_W$ for
proton, and $c_A\approx 1.25$ for both nucleons [\refcite{22,22a}].
We show that the relic neutralino can not overcome the threshold and
there are no detectable signals at all. In contrast, the high energy
neutralino overcome the threshold leading to a specific signature of
the final states, which is considered below in detail.

Firstly, let us consider the relic low-energy neutralino typical for
the (Cold) Dark Matter in the Galactic halo. In the case of pure
Higgsino states in the framework of our scenario, the lowest order
contributions to $\chi - N$ interaction correspond to the
spin-independent (SI) inelastic process [\refcite{17}]. It is a
consequence of the Majorana formalism, where all neutralino have the
same sign of masses, i.e. the same parity (see Ref.~[\refcite{17}]).
At the tree level, the cross-section for the relic
neutralino-nucleon $t$-channel reaction in the non-relativistic
limit mainly depends on the vector-like part of boson-nucleon
interaction, which leads to the SI contribution. By the
straightforward calculations one can get from Eq.~(\ref{3.a}) the
cross-section in the following threshold form:
\begin{equation}\label{3.1}
\sigma_{\chi
N}=\frac{G^2_F}{4\sqrt{2}\pi}\frac{c^2_V}{v_r}M^{3/2}_N(W_{\chi}\frac{M_N}{M_{\chi}}-\delta
m)^{1/2},
\end{equation}
where $v_r$ is the dimensionless relative velocity in units of the
speed of light $c$, $M_N$ is the nucleon mass, and the threshold
term $W_{\chi}(M_N/M_{\chi})-\delta m$ directly follows from the
kinematics of the process (see Appendix B for details). In the
non-relativistic case $W_{\chi}$ is an average kinetic energy of
neutralino in the Sun neighborhood, $W_{\chi}=M_{\chi}v^2_r/2$. For
$M_{\chi_1} \sim 1 \;\mathrm{TeV}$ this energy is $W_{\chi}\sim
1\;\mathrm{MeV}$. So, for $\delta m>1\,\mbox{KeV}$ the process is
forbidden and the relic neutralino cannot be detected in the direct
terrestrial experiments. The cross section for the
neutralino-nucleon scattering with the chargino production (so
called recharge process) is similar to Eq.~(\ref{3.1}), and for the
splitting $\delta m^- \approx 0.5\cdot\delta m>1\,\mbox{KeV}$ this
channel is also closed. Note that the phenomenology of the relic
neutralino-nucleon scattering crucially depends on the structure of
the $Z\chi\chi$-interactions. The absence of the diagonal vertex
(see Appendix A) $Z\chi_1\gamma_5\gamma_{\mu}\chi_1$ (spin-dependent
interaction) closes the direct channel of neutralino-matter
interaction. This is a principal consequence of the analysis which
have been made in Ref.~[\refcite{17}].

Now we consider the scattering of the non-relic neutralino off
nucleon above its threshold which is completely determined by the
mass splittings $\delta m$ and $\delta m^-$. As it was noted
earlier, these values are $\sim 10^2\,\mbox{MeV}$. The energy of
non-relic neutralinos can exceed the energy of the relic ones ($\approx
1\;\mbox{MeV}$) by many orders of magnitude. Such high-energy
neutralinos can be produced all the time due to the non-elastic
scattering of high-energy cosmic rays (or in various decay channels
of super-heavy $M_X$ particles). So, these high-energy neutralinos
do not directly connected with the Cold Dark Matter.

The structure of the neutralino-chargino-boson vertices (\ref{3.0})
allows s-channel transitions $\bar{f}f\to Z\to \chi_1\chi_2;
\bar{\tilde{H}}\tilde{H}$ and $f_u f_d\to W\to \chi_{1,2}\tilde{H}$.
These processes can take place in collisions of high-energy
particles from cosmic rays (or at the LHC), producing high-energy
$\chi_{1,2}$ and/or $\tilde{H}$. Further, $\chi_2,\,\tilde{H}$
intensively decay producing the energetic LSP, $\chi_1$. As a
result, some quantity of the lightest energetic LSPs, which are not
relic ones, can be accumulated and exists now.

Let us consider the interactions of these neutralinos with matter,
in particular, with nucleons. Obviously, the neutrino-nucleon
scattering provides the main contribution to the background. Due
to the fast decreasing of the cosmic rays density with the
increase of energy, we restrict ourselves to the processes with
the boson transverse momenta $Q^2\lesssim 2\,\mbox{GeV}$. This
restriction is consistent with the value of mass splitting and
makes it possible to evaluate maximal cross-section and to apply
the simplest vector and axial vector form-factors for the
description of $npW$ and $NNZ$ vertices ($N=n,p$). Amplitude of
the recharge process $\chi_1n\to\tilde{H}p$ is
\begin{equation}\label{3.4}
\mathit{M}=\frac{g^2}{4\sqrt{2}M^2_W}\bar{\chi}_1\gamma^{\mu}\tilde{H}\cdot
\bar{n}\gamma_{\mu}[c_V(Q^2)-c_A(Q^2)\gamma_5]p,
\end{equation}
where $c_V(0)=1$ and $c_A(0)\approx 1.25$. The $Q^2$-dependence for the
vector and axial-vector parameters can be chosen in the simplest
form $c_V(Q^2)=c_V(0)(1+Q^2/m^2_V)^{-2}$ and
$c_A(Q^2)=c_A(0)(1+Q^2/m^2_A)^{-2}$, where $m_V=0.84\,\mbox{GeV}$
and $m_A=0.90\,\mbox{GeV}$ in analogy with the neutrino-nucleon
scattering. In order to get a rough estimation of the cross-section,
it is sufficient to use these two form-factors only. In this
approximation, we get the cross-section of $\chi_1 n\to \tilde{H} p$
scattering in the form:
\begin{equation}\label{3.5}
\sigma(s)\approx\frac{G_F^2}{16\pi} f(s)
\int_{-1}^{1}[c^2_V(Q^2)+c^2_A(Q^2)]\,F(x)dx,
\end{equation}
where
\begin{align}\label{3.6}
f(s)&=\frac{s(1-M_{\tilde{H}}^2/s)}{[(1-M^2_{\chi}/s)^2-2M^2_N
M^2_{\chi}/s^2]^{1/2}},\,\,\,F(x,s)=b+\frac{1}{4}(a_1+bx)(a_2+bx)-\notag\\&\frac{M_{\chi}M_{\tilde{H}}}{s}b(1-x);\,\,\,
a_{1,2}=1\pm\frac{M^2_{\chi}-M^2_H}{s}-\frac{M^2_{\chi}M^2_{\tilde{H}}}{s^2};\,\,\,b=(1-\frac{M^2_{\chi}}{s})
(1-\frac{M^2_{\tilde{H}}}{s});\notag\\&Q^2=\frac{s}{2}(1+\frac{M^2_{\chi}-M^2_N}{s})(1+\frac{M_{\tilde{H}}^2-M^2_N}{s})
-\frac{s}{2}\bar{\lambda}(M^2_{\chi},M^2_N;s)\bar{\lambda}(M_{\tilde{H}}^2,M^2_N;s)x\notag\\-&M^2_{\chi}-
M_{\tilde{H}}^2;\,\,\,\bar{\lambda}(a,b;c)=(1-2\frac{a+b}{c}+\frac{(a-b)^2}{c^2})^{1/2};\,\,\,x=\cos
\theta.
\end{align}
where $\theta$ is the scattering angle. Note that the expressions
(\ref{3.5}) and (\ref{3.6}) are not applicable if $s\approx
M^2_{\chi_1}$, since our approximation ($M_N=0$ in the $F(x,s)$) is
violated in this regime. Note also that the cross-section in
relativistic case depends on vector ($c_V$) and axial-vector ($c_A$)
parts of the interaction. So, in this case the SI and SD
contributions are mixed. From the expression for the value $Q^2$ it
follows that our approach is restricted by neutralino energy
$E_{\chi}\lesssim 2M_{\chi}$ for the ``averaged'' value of the
scattering angle $\theta$. We have also derived the approximate
formulae for the neutralino threshold energy (in laboratory frame of
reference), $E^{thr}$, in the limit of small $M_N\ll M_{\chi}$:
\begin{align}\label{3.7}
&E^{thr}=(M_{\chi_1}+M_{\tilde{H}})\frac{\delta
m^-}{2M_N}+M_{\tilde{H}}.
\end{align}
From this relation it follows that in the case $M_{\chi_1}\sim
1\,\,\mbox{TeV}$ and $\delta m^{-}\sim 1\,\,\mbox{GeV}$ the
threshold energy is $E^{thr}\sim 2M_{\tilde{H}}$.


The typical value of the cross-section is about $\sigma\sim
3-7\,\,\,\mbox{fb}$ for $\delta m^- = 0.1\,\,\,\mbox{GeV}$ and
$E\sim 2 - 3\,\,\,\mbox{TeV}$. It decreases with increasing of the
mass splitting $\delta m^-$ and the threshold energy $E^{thr}$. As
it follows from the structure of $\chi_1\chi_2 Z$-vertex, formula
(\ref{3.5}) is also valid in the case of the neutral channel process
$\chi_1 N\to\chi_2 N^{'}$. So, the cross-section of this scattering
is also an order of few femtobarns at the considered energies. Thus,
to detect the events of the neutralino-nucleon scattering we need a
very massive and large-scale detector [\refcite{27}].

The cross-sections of the chargino-neutralino pair production at the
LHC in all permitted combinations ($\chi_1 \chi_2$, $\chi_1 \tilde
H$, $\chi_2 \tilde H$, $\tilde{H} \bar{\tilde{H}}$) were given in
Ref.~[\refcite{11}] in the framework of analogous ``low $\mu$''
Split SUSY model. For all processes the cross-sections are an order
of $10^{-3}- 10^{-5}\; \mathrm{pb}$ with $\mu = 800 - 1400
\;\mathrm{GeV}$. Therefore, further we discuss the possible
signatures of the final states only.

\section{Decay properties of the light chargino}

Now we turn to possible experimental signature of the
neutralino-nucleon scattering. To this end we are going to analyze
the decay channels of the products of scattering $\tilde{H}$ and
$\chi_2$. In particular, to study the recharge process $\chi_1 n\to
\tilde{H}p$ we have to deal with decays $\tilde{H}^-\to\chi_1
l^-\bar{\nu}_l,\,\chi_1\pi^-,\,\chi_1 \pi^{-}\pi^{0}$, where
$l=e,\mu$ and $\chi_1$ is LSP. The same decay modes define the final
states signature in production processes of $\tilde{H}$ at the LHC.
Note that the three-pion decay mode of $\tilde{H}$ is small for the
mass-splitting $\delta m^- \leq 1\,\mbox{GeV}$.

The branching ratios of these channels strongly depend on the mass
splitting $\delta m^-=M_{\tilde{H}}-M_{\chi_1}$. The $\tilde{H}$
decay rate in semi-leptonic channels has the following form
\begin{align}\label{3.9}
\Gamma_l=&\frac{G^2_F}{96\pi^3M_{\tilde{H}}}\int_{M_l^2}^{(\delta
m^{-})^2} dq^2
\bar{\lambda}(q^2,M_{\chi_1}^2;M_{\tilde{H}}^2)\bar{\lambda}(0,M_l^2;q^2)[q^2\bar{\lambda}^2
(0,M_l^2;q^2)
(M_{\tilde{H}}^2+M_{\chi_1}^2\notag\\-&4M_{\tilde{H}}M_{\chi_1}-q^2)+(1+\frac{M^2_l}{q^2}-2\frac{M^4_l}{q^4})
    ((M_{\tilde{H}}^2-M_{\chi_1}^2)^2-2M_{\tilde{H}}M_{\chi_1}q^2-q^4)],
\end{align}
where $\delta m^{-}=M_{\tilde{H}}-M_{\chi_1}$ and
$\bar{\lambda}(a,b;c)$ is normalized K\"allen function defined in
Eq.~(\ref{3.6}). Obviously, expression (\ref{3.9}) differs from the
analogous formula in Ref.~[\refcite{12-1,12-2}] for the case of
light gaugino, but numerical results of calculation are close. Note
also that analytical results coincides, if we rewrite a part of our
expression in other kinematical variables and cast away some terms,
which are subdominant numerically.

For $l=e$ we have $M_e\ll\delta m^-$, and the expression for the
corresponding decay width is simplified considerably
\begin{equation}\label{3.11}
\Gamma_e\approx\frac{G^2_F}{192\pi^3}(\delta m^-)^5.
\end{equation}
The decay rate of the process $\tilde{H}^-\to \chi_1\pi^-$ can be
calculated with the help of the well known soft pion matrix element
$<\pi|\bar{d}\gamma_{\mu}(1-\gamma_5)u|0>\;=f_{\pi}q_{\mu}/\sqrt{2q^0}$,
where $f_{\pi}\approx 132 \,\mbox{MeV}$ is the pion decay constant and $q$ is the
four-momentum of $\pi$-meson. Substitution of this equality into the
amplitude of the process $\tilde{H}\to \chi_1 \bar{u}d$ leads to the
decay rate in the one-pion channel
\begin{equation}\label{3.12}
\Gamma_{\pi}\simeq\frac{G^2_F}{4\pi}|U_{ud}|^2 f^2_{\pi}(\delta
m^-)^2
M_{\tilde{H}}\sqrt{1-2\frac{M^2_{\pi}+M^2_{\chi_1}}{M_{\tilde{H}}^2}+\frac{(M^2_{\pi}-M^2_{\chi_1})^2}{M_{\tilde{H}}^4}}.
\end{equation}
In the channel with two final pions the transition from quark to
hadron level is described by the matrix element [\refcite{31,31a}]
\begin{equation}\label{3.12a}
<\pi^-\pi^0|\bar{d}\gamma_{\mu}(1-\gamma_5)u|0>=\sqrt{2}F_{\pi}(q^2)(k_{-}-k_{+})_{\mu},
\end{equation}
where $q=k_{-}+k_{+}$ is the sum of $\pi^{-}$ and $\pi^{+}$ momenta.
Expression for $F_{\pi}(q^2)$ can be taken from
Refs.~[\refcite{31,32}]
\begin{align}\label{3.12b}
&F_{\pi}(q^2)=\frac{M^2_{\rho}}{M^2_{\rho}-q^2-iM_{\rho}\Gamma_{\rho}(q^2)}\exp\left\{-\frac{q^2\mathrm{Re}[A(q^2)]}
{96\pi^2f^2_{\pi}}\right\},\notag\\
&A(q^2)=\log\left(\frac{M^2_{\pi}}{M^2_{\rho}}\right)+8\frac{M^2_{\pi}}{q^2}-\frac{5}{3}+\sigma^3_{\pi}\log\left(\frac{\sigma_{\pi}+1}
{\sigma_{\pi}-1}\right),\,\,\,\sigma_{\pi}=\sqrt{1-4\frac{M^2_{\pi}}{q^2}}\notag\\
&\Gamma_{\rho}(q^2)=\theta(q^2-4M^2_{\pi})\frac{\sigma^3_{\pi}M_{\rho}q^2}{96\pi
F^2_{\pi}}\,,
\end{align}
where $F_\pi=f_\pi/\sqrt{2}\simeq$ 93 MeV (see
Refs.~[\refcite{31,32}]). With the help of Eqs.~(\ref{3.12a}) and
(\ref{3.12b}) we get the two-pion decay rate as
\begin{equation}\label{3.12c}
\Gamma_{2\pi}=\frac{G^2_F|U_{ud}|^2}{64\pi^3M_{\tilde{H}}}\int_{q^2_1}^{q^2_2}|F_{\pi}(q^2)|^2\sqrt{1-4\frac{M^2_{\pi}}{q^2}}f(q^2)
\bar{\lambda}(M_{\chi_1}^2,q^2;M_{\tilde{H}}^2)\,dq^2,
\end{equation}
where $q_1=2M_{\pi},\,\,q_2=\delta m^{-}$, $F_{\pi}(q^2)$ is defined
by Eq.~(\ref{3.12b}), and
\begin{align}\label{3.12d}
&f(q^2)=\frac{1}{6}(\delta
m^{-})^2(M_{\tilde{H}}+M_{\chi_1})^2-\frac{2}{3}M^2_{\pi} (\delta
m^{-})^2+\frac{8}{3}M^2_{\pi}M_{\tilde{H}} M_{\chi_1}\notag\\&+
q^2\left[\frac{1}{6}(\delta m^{-})^2-\frac{2}{3}M_{\tilde{H}}
M_{\chi_1}+\frac{4}{3}M^2_{\pi}\right]
-\frac{1}{3}q^4-\frac{2M^2_{\pi}}{3q^2}(\delta
m^{-})^2(M_{\tilde{H}}+M_{\chi_1})^2\,.
\end{align}
Making use of strong inequalities $M_{\pi}/M_{\tilde{H}}\ll
1,\,\delta m^-/M_{\tilde{H}}\ll 1$, the expression (\ref{3.12c}) can
be simplified
\begin{eqnarray}\label{2pi-simpl}
\Gamma_{2\pi}\simeq\frac{G^2_F|U_{ud}|^2}{48\pi^3}\int_{q^2_1}^{q^2_2}|F_{\pi}(q^2)|^2
\left(1-\frac{4M_{\pi}^2}{q^2}\right)^{3/2}((\delta
m^-)^2-q^2)^{3/2}\,dq^2
\end{eqnarray}
Analogous formulae were represented in Ref.~[\refcite{12-1,12-2}]
for the light gaugino decay. Changing the proper vertex functions,
we get close numerical results in our case. However, analytical
representations of the expressions for the decay rates are
different.

\begin{figure}[!h]
\centerline{\epsfig{file=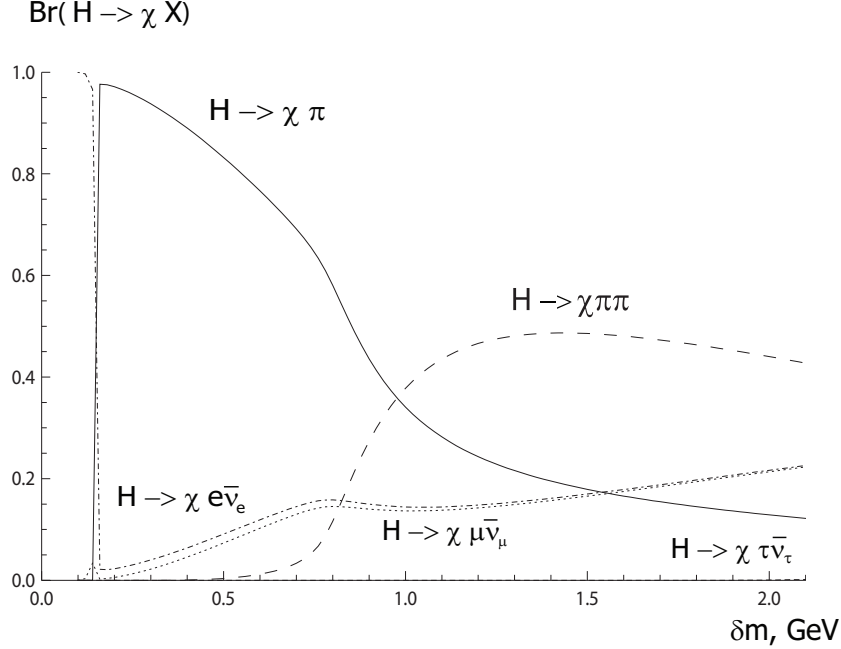,width=11cm}}
\caption{Branching ratios for chargino decays as functions of mass
splitting.} \label{fig:charge}
\end{figure}

Assuming $\Gamma^{tot}_H\approx
\sum_{l}\Gamma_l+\Gamma_{\pi}+\Gamma_{2\pi}$, we can evaluate the
branching ratios $B_l=\Gamma_l/\Gamma^{tot}_H$,
$B_{\pi}=\Gamma_{\pi}/\Gamma^{tot}_H$ and
$B_{2\pi}=\Gamma_{2\pi}/\Gamma^{tot}_H$, which describe the
signature of the total chargino decay process. Calculated ratios are
presented in Fig.~\ref{fig:charge} as functions of the mass
splitting $\delta m^-$ at fixed $M_{\tilde{H}}=1.4\,\mbox{TeV}$.

We see from Fig.~\ref{fig:charge} that the branching ratio of the
pion channel is strongly dominant above the threshold due to soft
nature of the pion production. With the increase of the mass
splitting $\delta m^-$, the processes go to the hard limit, and
$\mathrm{Br}(\pi)$ becomes an order of $\mathrm{Br}(l\bar{\nu}_l)$,
that is, an order of QCD limit $\mathrm{Br}(d\bar{u})$. The two-pion
decay mode is the most significant one at $\delta m^-\approx
1\mbox{GeV}$. Furthermore, we neglect the channels with production
of $K,\rho,...$ mesons in the final state. This is because the
channel with $K^{-}$-meson in the final state is suppressed by the
factor $|U_{ds}|^2\approx 0.05$, while decay into $\rho$ meson leads
to the two-pion final state. So, we simulate all hadronic decay
modes of chargino by pion final states. Certainly, in the QCD region
these reactions should be described by the amplitude $\tilde{H} \to
q' \bar q \chi_1$, as it was also considered quantitatively in
Ref.~[\refcite{13a}] in the SUSY scenario with $\mu\gg M_{1,2}$.

\section{Decay properties of the next lightest neutralino}

\begin{figure}[!h]
\centerline{\epsfig{file=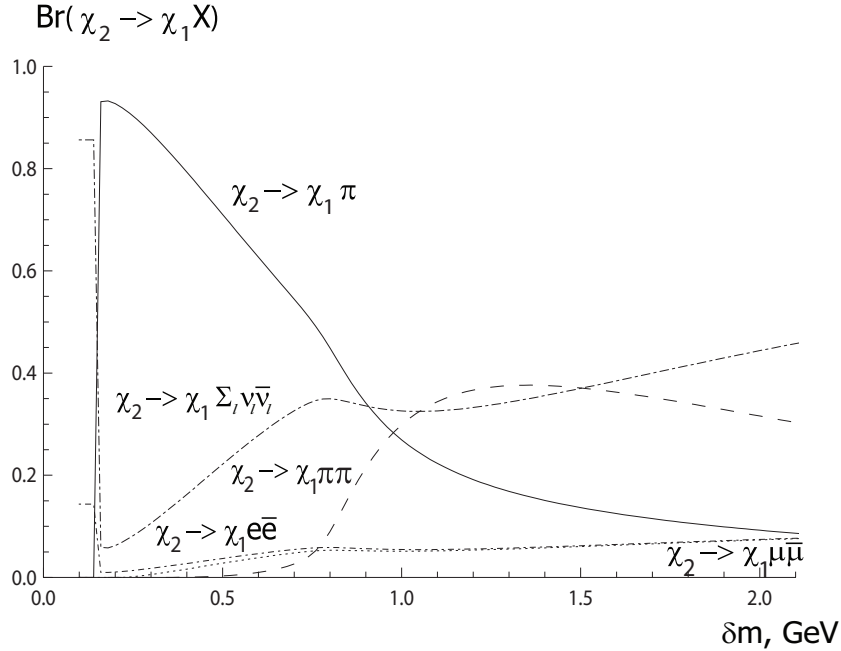,width=11cm}}
\caption{Branching ratios for neutralino decays as functions of mass
splitting.} \label{fig:neut}
\end{figure}

Now let us consider the neutral channel of the scattering $\chi_1
N\to\chi_2 N^{'}$ with the consequent decay of $\chi_2$. Because of
ambiguity concerning the neutralino mass splitting, we consider
$\delta m\sim \delta m^{-}$. As it was mentioned before, formula for
the cross-section is the same as for the recharge process. However,
signature of the total decay process is different. In this case, we
have the dominant decay channels $\chi_2\to\chi_1 f\bar{f}$, where
$f=e,\mu,\nu$ and $\chi_2\to\chi_1\pi^0\; \chi_2\to\chi_1\pi^+
\pi^-$. The semi-leptonic decay rate is
\begin{align}\label{3.13}
 \Gamma_l=&\frac{G^2_F}{64\pi^3M_{\chi_2}}\int_{4M_l^2}^{(\delta m)^2} dq^2
 \bar{\lambda}(q^2,M_{\chi_1}^2;M_{\chi_2}^2)\bar{\lambda}(M_l^2,M_l^2;q^2)\{\frac{c_{+}}{6}
 [q^2\bar{\lambda}^2 (M_l^2,M_l^2;q^2)\times\notag\\&\times(M_{\chi_2}^2+M_{\chi_1}^2-4M_{\chi_2}M_{\chi_1}-q^2)+(1+2\frac{M^2_l}{q^2})
    ((M_{\chi_2}^2-M_{\chi_1}^2)^2\notag\\-&2M_{\chi_2}M_{\chi_1}q^2-q^4)]
    + c_{-} M^2_l(M_{\chi_2}^2+M_{\chi_1}^2-4M_{\chi_2}M_{\chi_1}-q^2)\}
\end{align}
where $c_{\pm}=c^2_V\pm c^2_A$, and $\delta m=M_{\chi_2}-M_{\chi_1}
$. In particular, for the neutrino channel ($f=\nu$) we have
\begin{equation}\label{3.15}
\Gamma_{\nu}=\frac{G^2_F}{3\cdot256\pi^3}(\delta m)^5.
\end{equation}
The decay rate of the pion neutral channel is found in complete
analogy with the charge channel and reads
\begin{equation}\label{3.16}
\Gamma_{\pi}\simeq\frac{G^2_F}{8\pi} f^2_{\pi}(\delta m)^2
M_{\chi_2}\sqrt{1-2\frac{M^2_{\pi}+M^2_{\chi_1}}{M_{\chi_2}^2}+\frac{(M^2_{\pi}
-M^2_{\chi_1})^2}{M_{\chi_2}^4}}.
\end{equation}
And the corresponding decay rate of the two-pion neutral channel in
analogy with Eq.~(\ref{2pi-simpl}) is
\begin{equation}\label{3.16a}
\Gamma_{2\pi}\simeq\frac{G^2_F}{96\pi^3}\int_{q^2_1}^{q^2_2}|F_{\pi}(q^2)|^2
\left(1-\frac{4M_{\pi}^2}{q^2}\right)^{3/2}((\delta
m)^2-q^2)^{3/2}\,dq^2
\end{equation}
where $q_1=2M_{\pi},\,\,q_2=\delta m$ and $F_{\pi}(q^2)$ is defined
by Eq.~(\ref{3.12b}). Similarly to the charged case, for the neutral
scattering the signature is represented by the branching ratios of
the neutralino decay channels, $B_l=\Gamma_l/\Gamma^{tot}$,
$B_{\pi}=\Gamma_{\pi}/\Gamma^{tot}$ and
$B_{2\pi}=\Gamma_{2\pi}/\Gamma^{tot}$, which are represented in
Fig.~\ref{fig:neut}. These ratios are functions of the
mass-splitting $\delta m$ at fixed $M_{\chi_1}=1.4\,\mbox{TeV}$.

From Fig.~\ref{fig:neut} we conclude that behavior of the branching
ratios in the neutral case is similar to that for the chargino
decays. Namely, here again the process of pion production starts in
the soft regime and goes to the QCD limit (hard regime) with
increasing of the mass splitting $\delta m$. We should also mention that the decay channel with
$K^0$-meson is strongly suppressed due to absence of $Zds$ vertex at
the tree level. However, in analogy with the chargino decay, the
channels with $2\pi$ production is quite noticeable at $\delta m \approx 1\,\mbox{GeV}$.

\begin{figure}[!h]
\begin{minipage}{0.48\textwidth}
 \centerline{\includegraphics[width=1.05\textwidth]{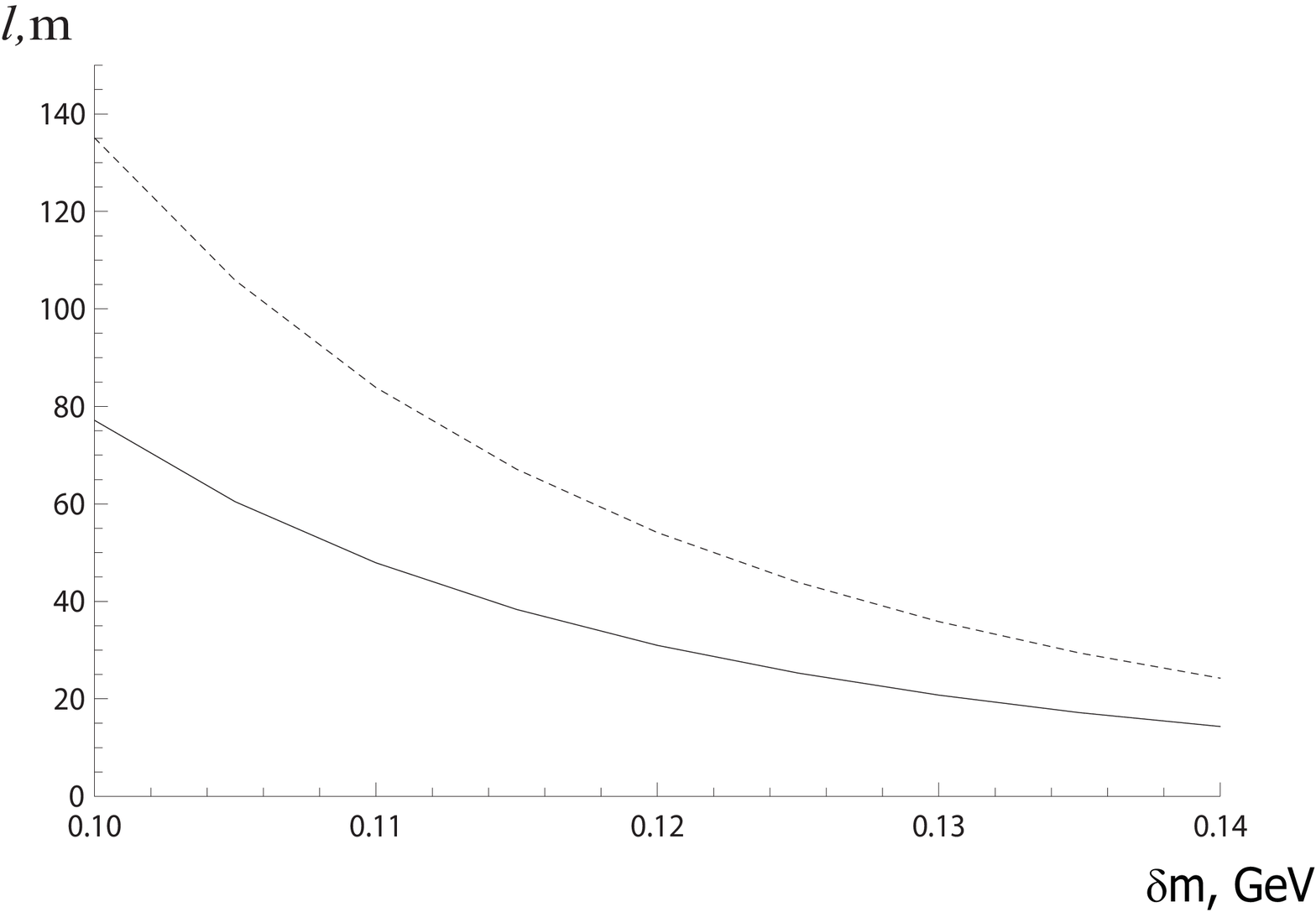}}
\end{minipage}
\begin{minipage}{0.48\textwidth}
 \centerline{\includegraphics[width=1.05\textwidth]{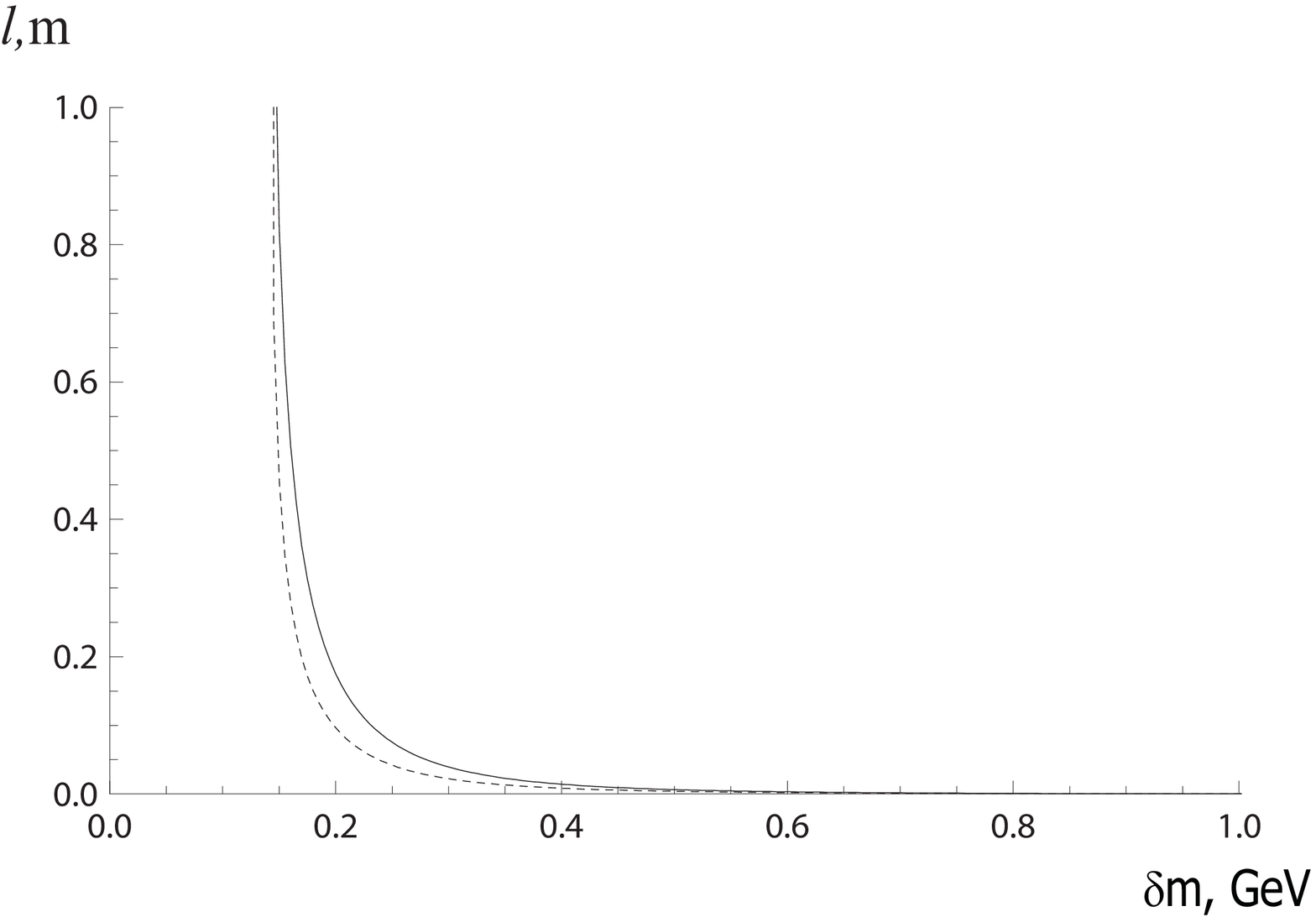}}
\end{minipage}
\caption{The range of neutralino (solid line) and chargino (dashed
line) as function of mass splitting.} \label{range}
\end{figure}

Collecting all results for the chargino and neutralino decay
channels in the scenario under consideration, we get the chargino
$\tilde{H}$ and neutralino $\chi_2$ ranges, $l=c\tau$ ($\tau$ is the
life time of the $\tilde{H}$ or $\chi_2$), depending on $\delta
m^{-}$ and $\delta m$, respectively. The corresponding curves are
shown in Fig.~\ref{range} (see, for comparison,
Refs.~[\refcite{12-1,12-2,13a}]). One can see that the chargino
track can be detected if the small mass splitting $\delta m^-$
occurs.

From the above presented results it follows that if $\delta m^-
\lesssim m_{\pi}$, we have no any visible signals from the chargino
decay $\tilde H\to l \nu \chi_1$ due to the final leptons softness.
Further, with the increasing of the mass splittings, when $m_{\pi} <
\delta m^-, \delta m \lesssim 1 \, \mathrm{GeV}$, chargino and the
NLSP decay mainly through one-pion and two-pion channels (see
Figs.~\ref{fig:charge} and ~\ref{fig:neut}). The charged pions can
be visible in experiment together with the chargino track. For the
NLSP with $\delta m \sim 1 \; \mathrm{GeV}$ the neutrino channels
are very important, i.e. fraction of the events with large missing
energy increases up to almost $50 \,\%$. In
Refs.~[\refcite{10,11,12-1,12-2,13a}], the analogous analysis of
possible final states was made for the case with the degenerated
lowest chargino and neutralino states. However, this analysis was
fulfilled for the SUSY scenario with the relatively light neutralino
and chargino states, $M_{\chi}\sim O(100 \;\mathrm{GeV})$. Moreover,
our results for the chargino and neutralino branching ratios are
different from the ones in papers mentioned above (cf.
Figs.~\ref{fig:charge} and \ref{fig:neut} and corresponding curves
from these papers).

It should be mentioned that charged Higgs bosons are arranged at
some high-energy scale, as it specifically takes place in Split SUSY
models (despite of one ``standard'' neutral Higgs boson). Then, all
additional contributions to the considered processes, which are
mediated by the heavy Higgs bosons, are suppressed. The same
suppression occurs in reactions with intermediate heavy squark (slepton) states.

\section{Conclusions}
The one-loop RG analysis of the SUSY $SU(5)$ theory leads to a few
sets of energy scales, which are compatible with the conventional ideas
on the DM structure and experimental expectations. Due to a specific
form of the RG equations the superscalar scale $M_0$ remains arbitrary, and
it occurs that the variety of possible scenarios can be divided into two
classes: $M_{1/2}\gg \mu$ or $M_{1/2}\ll \mu$.

In this paper, the hierarchy $M_0\gtrsim M_{1/2} \gg \mu$ (the Split
Higgsino model) and its possible experimental manifestations were
considered. For this case, the (one-loop) RG approach resulted to
the SUSY breaking scale $M_{SUSY} \sim 10^8 - 10^9\,\mathrm{GeV}$
and neutralino mass in the interval $1.2\,-\,1.6\,\mathrm{TeV}$.

We have considered the neutralino-nucleon scattering at low
(relic neutralino) and high energies. The high energy neutralino can
be produced as a result of high energy cosmic ray annihilation or in
decays of an exotic super-heavy $M_X$ particles. It was shown that
the energy of relic neutralino is far below the threshold in the
scenario under consideration. The scattering of high energy
neutralino off nucleon target can produce the nearest SUSY states,
$\chi_2$ and/or chargino $\tilde{H}$. We have analyzed in details the
signature of their decays, which make it possible to extract these events
from the background caused mainly by the neutrino-nucleon
scattering.

From the analysis we have fulfilled, it follows that in order to
detect some footsteps of Split Higgsino scenario, one have to
analyze the correlation of collider data, $\chi-N$ cross section
measurements and value of diffuse gamma flux from halo (or direct
photon spectrum). Only the comparison of all measured
characteristics could provide us with some conclusions on the
particular realization of the Split SUSY scenario. In a sense, this
model presents a class of ``Hidden SUSY'' scenarios which do not
reject SUSY ideas and, at the same time, can explain (possible)
absence of obvious SUSY signals at the LHC in the TeV region.

\vspace {0.5cm}

This work was supported in part by RFBR Grants No. 07-02-91557 and
No. 09-02-01149.
\appendix
\section{}
   Here we briefly analyze connection between the sign of neutralino masses and structure
   of neutralino-bozon interaction.
   The limit $M_Z/M_k \rightarrow 0$, where $M_k$ is $M_1, M_2$ or $\mu$,
   allows to simplify the analysis which can be used in the general case too.

If we omit the mixing of gauge and Higgs fermions, the mass
term of higgsino-like Majorana fields has the Dirac form:
\begin{equation}\label{A:2.1}
 M_h=\frac{1}{2}\mu(\bar{H}^0_{1R}H^0_{2L}+\bar{H}^0_{2R}H^0_{1L})
 +h.c.\,\,.
\end{equation}
This form can be represented by a ($2\times2$)\,- mass matrix having zero trace:
\begin{equation}\label{A:2.2}
    \mathbf{M}_2=
   \begin{pmatrix}
     0&\mu\\
     \mu&0
   \end{pmatrix}.
\end{equation}
There are two ways to diagonalize this matrix. The formal procedure
using the orthogonal matrix $\mathbf{O}_2$ leads to a spectrum with
opposite signs:
\begin{equation}\label{A:2.3}
 \mathbf{O}^T_2\mathbf{M}_2\mathbf{O}_2=\begin{pmatrix}
 \mu&0\\
 0&-\mu \end{pmatrix},\,\,\,\mathbf{O}_2=\frac{1}{\sqrt{2}}\begin{pmatrix}
 1&\,\,1\\
 1&-1 \end{pmatrix}\,,\,\,\,m_a=(\mu,-\mu)\,.
\end{equation}
 In this case, one of the Majorana fields has a negative mass
that is followed from the trace conservation
$\mathrm{Tr}\{\mathbf{O}^T_2\mathbf{M}_2\mathbf{O}_2\}=\mathrm{Tr}\{\mathbf{M}_2\}=0$.

The matrix $\mathbf{M}_2$ can also be diagonalized by the unitary
complex matrix $\mathbf{U}_2$, giving masses with the same sign
\begin{equation}\label{A:2.4}
 \mathbf{U}^T_2\mathbf{M}_2\mathbf{U}_2=\begin{pmatrix}
 \mu&0\\0&\mu\end{pmatrix},\,\,\,\mathbf{U}_2=\frac{1}{\sqrt{2}}
 \begin{pmatrix}1\,\,&i\\1&-i\end{pmatrix}\,,\,\,\,m_a=(\mu,\mu).
\end{equation}
Obviously, relation $H^0_L=(H^0_R)^C$ for the
Majorana spinor $H^0$ leads to the first term in (\ref{A:2.4}). This
relation defines procedure of diagonalization of
Majorana mass forms in the general case. The diagonalization (\ref{A:2.4}) is equivalent
to the procedure (\ref{A:2.3}) with the redefinition
$\chi\rightarrow i\gamma_5 \chi$ of the non-chiral (full)
field with $m=-\mu$ (for the chiral components
it corresponds to the transformation $\chi_{R,L}\rightarrow \pm i\chi_{R,L}$).

In this case, however, there is an infinite set of unitary matrices
$\mathbf{U}_{\phi}=\mathbf{U}_2\cdot\mathbf{O}_\phi$ which
diagonalize the mass matrix $\mathbf{M}_2$ (see [\refcite{17}] and
reference therein):
\begin{equation}\label{A:2.5}
 \mathbf{U}_{\phi}=\frac{1}{\sqrt{2}}\begin{pmatrix}e^{i\phi}&ie^{i\phi}\\
 e^{-i\phi}&-ie^{-i\phi}\end{pmatrix},\,\,\,\mathbf{O}_{\phi}=\begin{pmatrix}
 \cos \phi&-\sin \phi\\ \sin \phi&\cos \phi\end{pmatrix}.
\end{equation}
It has been shown in Ref.~[\refcite{17}] that the additional
$O_2$\,-\,symmetry leads to a free parameter arising in the general
case.

Dealing with the spinor field we should take into account the sign
of its mass in the propagator and polarization matrix or redefine
the field with a negative mass. As a rule this feature is not
considered in numerous phenomenological applications (see, [\refcite
{17}] and references therein). The redefinition of the Majorana
spinor $\chi^{'}=i\gamma_5 \chi $ changes the sign of mass in the
term $m\bar{\chi}\chi$ and does not change the term
$i\bar{\chi}\gamma^k\partial_k\chi$. This spinor transformation
saves the Majorana condition $(i\gamma_5\chi)^C=i\gamma_5\chi$ too.
Note also that for non-chiral Majorana field the redefinition
$\chi'=i\chi$  is not permissible. From the redefinition
$\chi'=i\gamma_5\chi$ it follows that the transformation properties
(relative to inversion) of Majorana fields having opposite mass
signs are different. It results to one usual Majorana field and one
pseudo-Majorana field. So, the mass signs are directly linked with
the relative parity, and this is important for the correct
interpretation of $Z\chi_1\chi_2$ interaction.

The gaugino mass subform is of the standard Majorana type and
the signs of the masses for $\chi_3$ and $\chi_4$ are defined by the signs of $M_1$
and $M_2$ in the case of small mixing. They can be made positive by the appropriate redefinition.

Let us consider for simplicity the pure higgsino approximation to
analyze the connection between the structure of boson-neutralino
interaction and the relative sign of neutralino masses. It can be
seen that the calculation rules should be different in two cases --
when masses of $\chi_2$, $\chi_1$ have opposite signs
(diagonalization (\ref{A:2.3})) and when they have the same signs
(diagonalization (\ref{A:2.4})).

The initial Lagrangian is
\begin{equation}\label{A:3.1}
 L_{int}=\frac{1}{2}g_Z
 Z_{\mu}(\bar{H}^0_{1L}\gamma^{\mu}H^0_{1L}+\bar{H}^0_{2R}\gamma^{\mu}H^0_{2R}).
\end{equation}
where $g_Z=g_2/\cos \theta_W$. The diagonalizations (\ref{A:2.3}) and (\ref{A:2.4}) lead to the following forms of neutralino-boson interactions, respectively:
\begin{equation}\label{A:3.2}
 (1)\; L_{int}=-\frac{1}{2}g_Z
 Z_{\mu}\bar{\chi}_2\gamma^\mu\gamma_5\chi^{'}_1;\qquad\quad
 (2)\; L_{int}=\frac{i}{2}g_Z Z_{\mu}\bar{\chi}_2\gamma^{\mu}\chi_1.
\end{equation}
In Eqs.~(\ref{A:3.2}) the first case, having opposite signs
$(\mu,-\mu)$, can be transformed into the second case with the same
signs $(\mu,\mu)$ by the redefinition $i\gamma_5\chi^{'}_1=\chi_1$.
It was shown in Ref.~[\refcite{17}] that both forms of $L_{int}$ in
Eqs.~(\ref{A:3.2}) give the same result without any field
redefinition if the negative sign of $\chi^{'}_1$ mass is taken into
account in calculations evidently. So, both structures in
Eqs.~(\ref{A:3.2}) lead to the parity-conserving vector interaction
giving the spin-independent contribution to the neutralino-nucleon
scattering. It is provided by the pseudo-Majorana nature of the
$\chi^{'}_1$ field.

As it has been shown in Ref.~[\refcite{17}], we cannot draw any
reasonable conclusions on the SD or SI contributions from the
interaction Lagrangian only, without consideration of the mass
signs. In other words, calculation rules should correlate with the
signs of neutralino masses. Specifically, the bilinear structures
$\bar{\chi}_2\gamma_{\mu}\chi_1$ and
$\bar{\chi}_2\gamma^{\mu}\gamma_5\chi^{'}_1$ are vectors, while
$\bar{\chi}_2\gamma_{\mu}\chi^{'}_1$ and
$\bar{\chi}_2\gamma^{\mu}\gamma_5\chi_1$ are axial vectors. The
structures $\bar{\chi}_2\chi^{'}_1$ and $\bar{\chi}_2\chi_1$ are
pseudoscalar and scalar, respectively. We conclude that the analysis
of the neutralino-nucleon interaction has to take into account
neutralino transformation properties. In particular, for the current
structure $\bar{\chi}_i\gamma^{\mu}\gamma_5\chi_k Z_{\mu}$ it is
possible to obtain SD or SI neutralino-nucleon cross sections
depending on the neutralino relative parity.

\section{}

In this Appendix we represent some technical details of calculations, which are important
to get the results above.

Let us consider the $\chi_1 N$ scattering in the non-relativistic
limit (see Fig.~1). The needed vertexes are described by the
Lagrangians:
\begin{equation}\label{B.1}
L_{Z\chi}=\frac{ig}{2\cos\theta_W}Z_{\mu}\bar{\chi}_2\gamma^{\mu}\chi_1,\,\,\,
L_{ZN}=\frac{ig}{4\cos\theta_W}Z_{\nu}\bar{N}\gamma^{\nu}(c_V-c_A\gamma_5)N.
\end{equation}
At $q^2\approx 0$ we use $c_V=1,\,\,\,c_A=1.25$ for the neutron and
$c_V=1-4\sin^2\theta_W,\,\,\,c_A=1.25$ for the proton. The process
$\chi_1 N\to\chi_2 N$ in the $t$-channel has an amplitude in
accordance with Eq.~(\ref{B.1}):
\begin{equation}\label{A.2}
\mathit{M}=\frac{ig^2}{8\cos^2\theta_W M^2_Z}\bar{\chi}_2(k_2)\gamma^{\mu}\chi_1(p_1)\cdot \bar{N}(p_2)\gamma_{\mu}(c_V-c_A\gamma_5)N(k_2),
\end{equation}
where $p_i$ and $k_i$ are the four-momenta of particles. Note, as it
was shown in Refs.~[\refcite{17a,22a}], in the non-relativistic case
the vector part of $ZNN$ vertex ($c_V$) gives spin-independent (SI)
contribution, while the axial-vector part of vertex ($c_A$) gives
spin-dependent contribution. Further we show that the vector type of
the $Z\chi_1\chi_2$ interaction cut off in the non-relativistic case
the vector part of $ZNN$ vertex and the cross-section in this case
depends on $c_V$ only. By the standard straightforward calculation
we get in the non-relativistic limit:
\begin{equation}\label{A.3}
|\mathit{M}|^2\approx K\{(c^2_V+c^2_A)[(p_1p_2)(k_1k_2)+(p_1k_2)(p_2k_1)]-2c^2_A M^2_N(p_1k_1)\},
\end{equation}
where $K$ is some numerical coefficient. Because the cross-section
value is proportional to the small momenta of $\chi_2$ and $N$ in
the CMS, we can use an approximation $(p_1 p_2)\approx
M_{\chi_1}M_N$, $(k_1k_2)\approx M_{\chi_2}M_N$ etc. in the
expression (\ref{A.3}). As a result, we have $|\mathit{M}|^2\approx
2K c^2_V M_{\chi_1}M_{\chi_2}M^2_N$, so the SI term only survives in
the case under consideration. Then, standard calculations give a
simple formula for the cross-section:
\begin{equation}\label{A.4}
\sigma\approx \frac{G^2_F}{8\pi v_r}c^2_V M_N k,
\end{equation}
where $k$ is absolute value of three-momenta of $\chi_2$ and $N$ in
the SCI, which is usually defined by the K\"allen function
\begin{equation}\label{A.5}
k=\frac{\sqrt{s}}{2}(1-2\frac{M^2_N+M^2_1}{s}+\frac{(M^2_N-M^2_2)^2}{s^2})^{1/2}.
\end{equation}
Here $s=(p_1+p_2)^2=(k_1+k_2)^2$ and $M_{1,2}=M_{\chi_{1,2}}$. The
value of $k$ can be found from the expressions for $s$ in the
laboratory coordinate system (LCS) and center-of-mass system (CMS):
\begin{align}\label{A.6}
s&=(p_1+p_2)^2\approx (M_1+M_N)^2+M_1M_N v^2_r,\,\,\,\mbox{(LCS});\notag\\
s&=(k_1+K_2)^2\approx
(M_2+M_N)^2+k^2\frac{M_2}{M_N},\,\,\,\mbox{(CMS)}.
\end{align}
From these equations it follows:
\begin{equation}\label{A.7}
k=\sqrt{\frac{M_N}{M_2}}\{M_1 M_N v^2_r-2\delta m (M_1+M_N)\}\approx\sqrt{2M_N}(W_1\frac{M_N}{M_1}-\delta m)^{1/2},
\end{equation}
where $\delta m=M_2-M_1$ and $W_1=M_1 v^2_r/2$ is the kinetic energy
of neutralino $\chi_1$ in LCS ($v_r$ is the dimensionless relative
velocity in units of the speed of light $c$). Finally, we get
formula for the non-relativistic neutralino-nucleon scattering:
\begin{equation}
\sigma=\frac{G^2_F}{4\sqrt{2}\pi} \frac{c^2_V
}{v_r}M^{3/2}_N(W_{\chi}\frac{M_N}{M_{\chi}}-\delta m)^{1/2}.
\end{equation}
Thus, the neutralino-nucleon scattering has two features --- the
main contribution is spin-independent and the threshold of the
reaction is $W^{thr}_{\chi}=\delta m\cdot (M_{\chi}/M_N)$.

\end{document}